\documentstyle[epsf,twocolumn,floats,prb,aps]{revtex}    
\newcommand{\figwidth}{3.375in}     
\begin{document}
\draft

\twocolumn[\hsize\textwidth\columnwidth\hsize\csname @twocolumnfalse\endcsname

\title{
Path integral Monte Carlo simulation of the second layer of
$^4$He adsorbed on graphite
}
\author{ Marlon Pierce and Efstratios Manousakis}
\address{
Department of Physics and Center for Materials Research and Technology,
Florida State University, Tallahassee, FL 32306-4350
}
\date{\today}
\maketitle
\begin{abstract}
\noindent
We have developed a path integral Monte Carlo method for simulating
helium films and apply it to the second layer of helium adsorbed
on graphite.  
We use helium-helium and helium-graphite interactions that are found
from potentials which realistically
describe the interatomic interactions.  The Monte Carlo sampling is
over both particle positions and permutations of particle labels.
From the particle configurations and static structure factor
calculations, we find that this layer possesses, in order of increasing
density, a superfluid liquid phase, 
a $\sqrt 7 \times \sqrt 7$ commensurate 
solid phase that is registered with respect to the first layer, 
and an incommensurate solid phases.  
By applying the Maxwell construction to the dependence of the low-temperature
total energy on the coverage, we are able to identify coexistence regions 
between the phases.  From these, we deduce an effectively zero-temperature
phase diagram.  Our phase boundaries are in agreement with heat capacity and 
torsional oscillator measurements, and demonstrate that the experimentally
observed disruption of the superfluid phase is caused by the growth
of the commensurate phase.
We further observe that the superfluid phase has a transition
temperature consistent with the two-dimensional value.  
Promotion to the third layer occurs for densities 
above 0.212 atom/$\AA^2$, in good agreement with experiment.  
Finally, we calculate the specific heat for each phase and obtain
peaks at temperatures in general agreement with experiment.  
\end{abstract}

]

\section{Introduction}
\label{sec:intro}

Helium adsorbed on graphite provides 
an excellent realization of a number of nearly two-dimensional 
(2D) phenomena.
The helium film grows in a succession of distinct, atomically thin layers
as the density of the adsorbate increases,       
and as many as seven such layers may be observed on a well-prepared 
substrate.\cite{zimmerli92}  Consequently, it is possible to 
investigate the evolution of each layer's phase diagram. 
A number of experimental methods have been used for this purpose, including  
specific heat measurements,
\cite{dash78,schick80,greywall91,greywall93,zimmerli92}
neutron scattering,\cite{nielsen80,carneiro81,lauter91,lauter92}
torsional oscillator measurements,\cite{reppy93,reppy96} and 
third sound.\cite{zimmerli92}  The phase diagrams of the layers nearest
the substrate are rich.  Evidence has been found for self-bound fluid 
phases that are superfluid at low temperatures, a variety of 
registered solid structures, and incommensurate solid phases.
These phases and the coexistence regions that separate
them are governed by a delicate balance of quantum effects, 
such as large zero-point motion and particle permutations, with
adatom and substrate interactions.

Much of the early experimental work on the helium-graphite system
concentrated on the first adsorbed layer.  Several reviews of this
work are available.\cite{dash78,schick80,bruch97}
On the other hand, until recently, relatively 
little information was available on the phases of the second and 
higher layers.  
This situation has changed dramatically over the last several years.
Extensive heat capacity measurements\cite{greywall91,greywall93} 
of the first six layers have been
performed, and superfluidity in the higher layers has been detected
by both torsional oscillator\cite{reppy93,reppy96} and third 
sound measurements.\cite{zimmerli92}  Taken together, these experiments
indicate that the second layer has a unique phase diagram,
with superfluid, commensurate solid, and
incommensurate solid phases.  No other layer exhibits all three phases. 

Motivated by these experiments, we have undertaken a path integral
Monte Carlo (PIMC) simulation of the second adsorbed layer.  
We identify a liquid (L) phase with an equilibrium density 
of 0.1750 atom/$\AA^2$, a $\sqrt 7 \times\sqrt 7$ commensurate 
triangular solid (C) at 
0.1996 atom/$\AA^2$, and an incommensurate triangular 
solid (IC) phase for densities above
0.2083 atom/$\AA^2$.  All coverage values are for the total adsorbed
film.  Using the Maxwell construction, we determine coexistence 
regions between these phases, namely the gas-liquid (G-L), 
liquid-commensurate solid (L-C), and commensurate-incommensurate 
solid (C-IC) phases, 
at effectively zero temperature.  Our calculated phase diagram 
confirms the idea that
the superfluid phase is interrupted by the formation of the 
commensurate solid.\cite{reppy93,reppy96,greywall93}
We further show that the liquid phase behaves like a typical two-dimensional
superfluid.  We also calculate the specific heat for each phase and
find peaks in general agreement with the experimental values.
Finally, we observe promotion to the third layer at a coverage
in good agreement with experiment.
A preliminary report of some of our findings has been 
published elsewhere.\cite{pierce98}
The present paper expands and extends this work.

This paper is arranged in the following manner.  Section \ref{sec:revexp}
provides an overview of what is known about the
second layer from experiments.  In Sec. \ref{sec:revsim}, we review
previous simulations of helium films and the related simulation of
two-dimensional helium.  We point out that none of these simulations,
while interesting in their own right, exhibit all the
phenomena observed in the second-layer phase diagram.  
Section \ref{sec:method} presents the details of our 
simulation method, which includes particle permutations and
realistic particle-particle and particle-substrate interactions.  The results
of our calculations are presented in Sec. \ref{sec:results}.  We 
demonstrate the existence of each phase, explain the construction
of the second-layer phase diagram, and present calculations of properties for
each phase.  

\subsection{Experimental Overview}
\label{sec:revexp}

Specific heat measurements have formed the basis for constructing the 
first helium layer's phase diagram, but until recently,
relatively little work was done on the second layer, with a couple of
exceptions.  Bretz\cite{bretz73}
examined this layer under compression of the third and obtained 
evidence for the melting of the incommensurate second
layer solid.  The low density range of this  
layer was explored by Polanco and Bretz.\cite{bretz78}
They determined that the compression of
the first layer by the growth of the second
must be taken into account in order to determine the phases
at low second-layer densities.  They interpreted their 
results to indicate that the second
layer has gas-liquid coexistence at low coverages.  

The heat capacity measurements of Greywall and Busch 
provide the most extensive investigation of the second-layer phase 
diagram.  They
find evidence for four phases:  gas, liquid, commensurate
solid, and incommensurate solid.   These phases are identified in
the following manner.
At low densities, a low, rounded peak
occurs in the heat capacity.  This has previously been associated with
the liquid phase.\cite{bretz78}  At low temperatures, the heat capacity
depends linearly on density roughly between 0.13 and 0.16 atom/$\AA^2$,
which is a requirement for coexisting phases,\cite{dash75}  Thus this
region can be identified as a gas-liquid coexistence region, with
the gas phase having negligible density at the lowest temperatures.  Evidence
for liquid-commensurate solid coexistence can be found between 0.187 and
0.197 atom/$\AA^2$.  In this region, in addition to the low peak associated
with the liquid phase, another, larger peak at a higher temperature can 
be observed.  The location of 
the larger peak is independent of coverage,
suggesting that it may be associated with the melting of a commensurate
solid phase.  Greywall suggested \cite{greywall93} 
that this phase corresponds to the $\sqrt 7 \times \sqrt 7$ commensurate
structure proposed earlier for $^3$He on graphite.\cite{elser89,abraham90}
A third coexistence region occurs between 0.2030 and 
0.2080 atom/$\AA^2$, where the commensurate melting peak is accompanied by
another, lower temperature peak.  This second peak is associated with
the melting of an incommensurate solid phase.  For coverages from 0.2080 to
the beginning of third layer promotion at 0.212 atom/$\AA^2$, the 
incommensurate melting peak 
is the sole feature in the specific heat measurements.
Unlike the peak associated with the commensurate
phase, the incommensurate melting peak is temperature dependent, occurring
at about 1 K at the lowest incommensurate densities, but increasing to about
1.5 K at the density where third layer promotion begins.

The principal limitation on using the heat capacity measurements 
to determine the phase diagram is that they
can only identify phases indirectly, so additional confirmation is desirable.
Direct evidence for the incommensurate solid phase
comes from neutron scattering,\cite{carneiro81,lauter91,lauter92} but
no similar evidence exists for the commensurate phase.  Apparently,
the incommensurate phase can be resolved in these experiments 
only after some additional compression
by the third layer.  Consequently, there is no scattering evidence for
the commensurate solid, which is replaced by the incommensurate solid 
before promotion to the third layer begins.  

Further insight into the second-layer phase diagram comes from the torsional
oscillator measurements of Crowell and Reppy.\cite{reppy93,reppy96}
They detected superfluidity at intermediate densities, which incidentally
provided direct evidence 
that the second layer has a liquid phase.  Questions remain
about the liquid phase, however, since the apparent onset density
is somewhat higher than would be expected from either the heat capacity 
measurements or the liquid equilibrium density of purely two-dimensional 
helium.\cite{whitlock88}  The range of superfluid coverage also provides 
additional, although indirect, evidence that a solid phase begins to 
form above 0.187 atom/$\AA^2$.  Above this density, the superfluid 
signal vanishes and does not reappear until the third layer.
This disappearance coincides almost
exactly with the growth of the supposed commensurate solid phase.  Apparently,
the growing solid phase disrupts the connectivity required to detect
superfluidity.

\subsection{Previous Simulations}
\label{sec:revsim}

The results of Monte Carlo calculations are often used to help interpret
the experimental results discussed above.  The simplest way to treat
a helium layer is as a purely two-dimensional system, 
for which there are
both zero temperature and finite temperature calculations.  
Whitlock et al. \cite{whitlock88}
used Green's function Monte Carlo to calculate the equilibrium liquid
coverage at zero temperature, obtaining 0.04356 atom/$\AA^2$.
They also determined that 2D helium would solidify, and that
the liquid and solid phases coexisted between 0.0678 and 0.0721 atom/$\AA^2$.
More recently, Gordillo and Ceperley\cite{gordillo98}
have investigated the 2D phase diagram at finite temperatures with path
integral Monte Carlo.  
Their low temperature results are consistent with the zero temperature
calculations.  
They also determined spinodal lines and found a finite density gas phase
at temperatures above 0.75 K.
The direct comparison of these results with the second helium 
layer is limited, since the 2D calculations do not include any substrate  
features and do not allow the film to spread perpendicularly.  As a 
result, no commensurate solid phase or layer promotion can occur.

Simulations of helium films using realistic models
for the graphite substrate have also been
performed.  Abraham and Broughton\cite{abraham87} used path integral
Monte Carlo to investigate the first layer of $^3$He on graphite.  They
were able to identify fluid, commensurate solid, domain wall liquid and solid, 
and incommensurate
solid phases.  Notably, they determined that particle permutations were
unimportant for the first layer for the coverages they investigated, so
there was no possibility for superfluidity in the simulation.
Experimentally, the phase diagrams for
$^3$He and $^4$He at the densities they simulated are nearly identical,
so it reasonable to conclude that their simulation results also apply
to $^4$He.  This work was extended\cite{abraham90} to a simulation of 
the second adsorbed layer
of $^3$He at the $\sqrt 7 \times \sqrt 7$ commensurate density.  Particle
permutations were again neglected.  It was established that the second-layer
commensurate phase was stable for temperatures below 1 K.  Very recently,  
Whitlock et al.\cite{whitlock98} investigated the ground state properties
of the first helium layer using a laterally averaged potential
for the helium-graphite interaction.  They determined the equilibrium liquid
coverage and the onset coverage for solidification in the first layer, and
determined the coexistence region between these two phases.  They 
also estimated completion densities for the first and second layers, obtaining
agreement with the experimental results.  They did not take the corrugations
of the graphite substrate into account and so did not observe the 
$\sqrt 3 \times \sqrt 3$ commensurate solid phase that occurs in the
first layer.  

Complementary to the calculations discussed above is the work 
by Clements
et al. \cite{clements93a,clements93b,saslow96} using the hypernetted-chain 
Euler-Lagrange theory.  
For 2D helium, this method 
reproduces the Monte Carlo results\cite{whitlock88} for the liquid phase
and provides a direct calculation of the chemical potential, third sound, and
spinodal points.  When applied to layered systems, 
the theory gives liquid coverage ranges and layering transitions but is not
capable of investigating solid phases.  For this reason, these
calculations are restricted to the
third and higher helium layers, and assume that the first two 
layers form an inert, featureless solid.
Also complementary are the path integral Monte Carlo 
calculations of Wagner and Ceperley\cite{wagner94,wagner96} 
for $^4$He and hydrogen films on crystalline hydrogen.  In their
helium film simulation, superfluidity and layer-by-layer growth
occurred, but the film did not solidify.  

As we discussed in Sec. \ref{sec:revexp}, 
the second layer of $^4$He on graphite is unusual in that it is known
experimentally to have both a superfluid liquid and two solid phases, one
commensurate and the other incommensurate with the first layer. 
The simulations discussed above are interesting in their own right, but
none have exhibited the three phases seen in the second layer.  
In order for a simulation to produce these phases, it must possess three
features.  First, the presence of superfluidity means that particle 
permutations must be included in the simulation.  This is
because superfluidity results from permutation cycles of infinite 
length.\cite{feynman53}  It is also expected
that the boundaries of the phases will be effected by permutations.
Second, the commensurate second-layer solid is found to be registered with
respect to the first layer, so the effect of first-layer atoms
must be taken into account.  Third, the attraction of the 
substrate and first layer on the second must be implemented correctly so that
the commensurate phase is replaced by the incommensurate phase
before promotion to the third layer begins.  
In the following 
section we outline our simulation method, which contains the necessary 
features to exhibit these three phases.  


\section{Simulation methods and details}
\label{sec:method}

Path integral Monte Carlo is a powerful tool for simulating quantum
systems at finite temperatures.  By incorporating sampling of
particle configurations and
particle permutations, both normal and superfluid
helium can be simulated.\cite{ceprev}  If a substrate is added
to the simulation, a quantum film will result.  The purpose of
this section is to describe the modifications that are necessary to add 
the effects of the substrate into the simulation.  The result will be
a simulation method that is capable of exhibiting superfluid, commensurate
solid, and incommensurate solid phases, as well as layer promotion.

Central to our PIMC method is the approximation used for the 
high temperature density matrix.  It is essential that the starting
temperature be made as low as possible so that permutations will
be accepted.  As we will discuss in this section,
the graphite substrate complicates a straight-forward extension
of the starting approximation used in bulk simulations.  For this reason
we will not include sampling of the first-layer atom configurations
in the calculation and will concentrate instead on the second layer.

It is essential to include the effect of the first layer on the second,
however.  We approximate this effect by placing first-layer atoms 
on the sites of a triangular lattice at a fixed height above the substrate.
This allows us to 
treat the helium-graphite correlations
in a much simpler manner, since the atoms on the second layer are not 
effected by the corrugations of the graphite substrate.  By not sampling 
first-layer configurations, we are also able to increase the number 
of second-layer atoms in the simulation.  In turn, this
allows us to scan second-layer coverages in a sufficiently
fine grid to observe coexistence regions.  Having a fine grid
is particularly important
for high second-layer densities, since the liquid-commensurate solid and 
commensurate-incommensurate solid coexistence 
regions exist over relatively narrow ranges.  

The trade-off for 
using this approach is that we 
ignore zero-point motion in the first
layer.  This will cause the second layer to form closer
to the first layer and have a narrower density profile.\cite{novaco76}
Ignoring the response of the first layer to the second 
is also known to lead to 
a lowering of the energy of a layer of helium adsorbed onto solid 
hydrogen.\cite{wagner94}
However, experimental results indicate
that neglecting zero-point motion in the first layer of helium on graphite
atoms is a reasonable approximation.  First, the 
Debye temperature of the solid first layer is greater than 
50 K, and it may be treated as a 2D Debye solid
up to 3 K.\cite{dash83}  In our simulation, the temperature is as
low as 200 mK, and never exceeds 2.2 K, so the first layer is 
relatively stiff. 
Second, although the first layer is known to be compressed by the growing
second-layer, this is most important at low second-layer 
densities, just after second-layer promotion begins.\cite{bretz78}  The
coverages studied by Polanco and Bretz\cite{bretz78} are below the range of our
simulation.  As we shall see, our approach is sufficient to
reproduce many of the observed features of the second layer.

\subsection{Path integral representation of the partition function}
We wish to study the problem of a quantum N-particle system in the presence
of a substrate.  The Hamiltonian for this system may be written as
\begin{eqnarray}
\label{eq:hamiltonian}
H = &-&\hbar^2/2m \sum_{i=1}^{N} {\bf\nabla}_i^2 + \sum_{i<j}^N v_{2B}(|{\bf r}_i - {\bf r}_j|) \nonumber \\
	&+& \sum_{i=1}^N v_{sub}({\bf r}_i),
\end{eqnarray}
where $v_{2B}$ is the spherically symmetric two-body potential between 
particles, and $v_{sub}$ is
the external field produced by the substrate.  The two-body potential for
helium is accurately represented by the 
Aziz potential.\cite{aziz92}  Previous path integral simulations
using this potential have proven
quite capable of reproducing numerous properties of 
liquid helium.\cite{cep86,cep87,runge92,ceprev}  The potential between
helium and graphite has been investigated by
Carlos and Cole.\cite{cole80}  Using helium-scattering data, they 
evaluated several forms for the helium-graphite potential.  
In order to write this potential in a pair form, anisotropic terms that
effectively enhance corrugation must be included.  Of the potentials
examined, an 
anisotropic 6-12 Lennard-Jones potential was found to be preferable,
although the form was not uniquely determined.
For helium atoms more than 4 $\AA$ above the
substrate, corrugations are negligible, and the anisotropic potential can be 
replaced by a laterally averaged potential that depends only on the
height of the atom above the substrate.   

The statistical mechanics of quantum systems are governed by 
the density matrix 
and the partition function.  For a system of $N$ bosons at an inverse
temperature $\beta$, the density matrix is given by
\begin{eqnarray}
\label{eq:denmat}
\rho({\bf R}, {\bf R}';\beta)=
\frac{1}{N!} \sum_P < {\bf R} |e^{-\beta H} | P {\bf R}'>,
\end{eqnarray}
where $\bf R$ and $\bf R'$ are two configurations of $N$ bosons.
The sum over $P$ is over all permutations of particle 
labels, and $P\bf R'$ is one such permutation.  Permutations lead
directly to the off-diagonal long-range order that produces superfluidity.
The partition function, $Z$,
is found by integrating the diagonal elements of the density matrix,
\begin{equation}
\label{eq:partit}
Z=\frac{1}{N!} \sum_P \int \rho({\bf R},P{\bf R},\beta) d^3 R.
\end{equation}

Evaluating the partition function for interacting systems at very
low temperatures is complicated by the fact that the 
kinetic and potential terms in the exponent of the density matrix cannot be 
separated, so the form of the density matrix is not known in,
for instance, the configuration space representation.  We can avoid this
problem by inserting $M-1$ intermediate configurations into
Eq. (\ref{eq:partit}) to obtain the path integral formulation of 
the partition function,
\begin{eqnarray}
\label{eq:pathint}
Z \ = & & \frac{1}{N!} 
	\sum_P \int...\int d^3 R_1 ... d^3 R_{M-1} d^3 R \nonumber \\
  & & \times \rho ({\bf R},{\bf R}_1; \tau) \rho({\bf R}_1, {\bf R}_2; \tau) 
  	\ldots \rho ({\bf R}_{M-1},P {\bf R}; \tau), 
\end{eqnarray}
where $\tau=\beta/M$.  The problem of evaluating the partition function
at a low temperature, $\beta^{-1}$, has been replaced by the problem 
of multiple integrations of density matrices at a higher 
temperature, $\tau^{-1}$.  The advantage of this is that the
high temperature density matrices may be approximated. 
In practice, the integrals appearing 
in Eq. (\ref{eq:pathint}) cannot
be directly evaluated for systems of strongly interacting 
particles.  Monte Carlo 
sampling may be used instead to generate
configurations and calculate observables.

Equation (\ref{eq:pathint}) lends 
itself to an interesting visualization.
The $N$ quantum particles can be thought of as $N$ interacting classical
ring polymers, each with $M$ beads.  Sampling the partition function then
corresponds to sampling the possible configurations of these polymers.  
Furthermore, particle permutations may be introduced into the Monte Carlo
method by splicing together two or more polymer chains.  
This is known as the polymer isomorphism.  

\subsection{Approximating the density matrix}
In order to use Monte Carlo sampling on the partition function, we must first
provide a starting approximation for the high temperature density matrices
that appear in the integrand of Eq. (\ref{eq:pathint}).  The simplest 
starting approximation is to use a very large $M$, which allows us to
separate the density matrix into kinetic and potential energy terms.
This is the semiclassical
approximation and is exact in the limit 
$M \rightarrow \infty$, according to the Trotter theorem.
For superfluid helium systems it is necessary to go beyond the semiclassical
approximation so that the starting temperature may be made as low as possible.
This makes sampling the permutations feasible and speeds the 
equilibration of the ring polymers by avoiding excessively long chains.
The high-temperature density 
matrix we introduce below can be used with starting temperatures as
low as 40 K.  We thus only have to use, for instance, $M=40$ to simulate
a system at 1 K.  

We approximate the high temperature density matrix as a product of the
exact free particle solution, an effective two-body interaction found from
the exact solution for two interacting helium atoms, and an effective 
external interaction found from the exact solution for a single atom
in a graphite potential:
\begin{eqnarray}
\label{eq:hitemp}
\rho({\bf R},{\bf R}';\tau) \ \approx
	&&\prod_{i=1}^{N} \rho^{free}_1({\bf r}_i,{\bf r'}_i;\tau) \nonumber \\
	&\times& \prod_{i=1}^{N} \tilde{\rho}^{Gr}_1({\bf r}_i,{\bf r'}_i;\tau)\nonumber \\
	&\times& \prod_{i<j}^{N} \tilde{\rho}^{He}_2({\bf r}_{ij}, {\bf r'}_{i,j};\tau),
\end{eqnarray}
where ${\bf r}_{ij}={\bf r}_i -{\bf r}_j$.  The terms $\rho^{free}$,
$\tilde{\rho}^{He}_1$, and $\tilde{\rho}^{Gr}_2$ will be discussed below.
This approximation assumes that
three-body contributions are negligible and that the helium-helium and
helium-graphite interactions can be decoupled.  The former has been shown
to be valid for bulk helium systems with starting temperatures as low as
40 K.  

The term $\rho_1^{free}$ is the 
density matrix for a free particle of mass {\em m}, given by
\begin{eqnarray}
\label{eq:free3d}
\rho^{free}({\bf r},{\bf r}';\tau) = \lambda_t^{-3} 
		\exp[-\pi({\bf r}-{\bf r'})^2/\lambda_t^2].
\end{eqnarray}
where $\lambda_t = \sqrt{2\pi\tau\hbar^2/m}$ is the mean thermal wavelength
for the temperature $1/\tau$. 

The helium-helium term, $\tilde{\rho}^{He}_2$, is the interacting part of the
solution to the density matrix for two helium atoms.  This can be found 
by separating the density matrix into
center-of-mass and relative components.  The density matrix for the relative
coordinates is a solution to 
\begin{eqnarray}
\label{eq:bloch_he}
\frac{\partial \rho^{He}}{\partial \tau}({\bf r}_{ij},{\bf r'}_{ij};\tau)
	=&[&(\hbar^2/m) \nabla^2 \nonumber \\
	&-& V^{He}({\bf r}_{ij})]\rho^{He}({\bf r}_{ij},{\bf r'}_{ij};\tau).
\end{eqnarray}
This equation is equivalent to that satisfied by the time 
evolution propagator in imaginary time.  
We solve this equation using the methods discussed by 
Ceperley.\cite{ceprev}  Briefly, the density matrix can be expanded 
in a series of partial waves and the expansion coefficients are
found by using the matrix-squaring
method.  The resulting solution is used to define the effective
helium-helium interaction, 
$U^{He}({\bf r}_{ij},{\bf r'}_{ij};\tau) \equiv -\ln(\tilde{\rho}^{He})$
where $\tilde{\rho}^{He}=\rho^{He}/\rho^{free}$.
This is a six-dimensional function, but the 
spherical symmetry of the density matrix allows us to approximate it as
a series of one-dimensional functions.  
This greatly reduces the memory requirements and increases
the speed at which the density matrix can be evaluated for a particular
configuration.  

The density matrix for a single helium atom above a graphite
substrate is a solution to
\begin{equation}
\label{eq:bloch}
\frac{\partial \rho^{Gr}_1}{\partial \tau}({\bf r},{\bf r'};\tau)
	=[(\hbar^2/2m) \nabla^2 - V^{Gr}({\bf r})]\rho^{Gr}_1({\bf r},{\bf r'};\tau),
\end{equation}
where  $V^{Gr}({\bf r})$ is the full graphite potential.
The helium-graphite term, $\tilde{\rho}^{Gr}_1$, is the
interacting part of the solution to this equation.
Near the substrate, the 
potential $V^{Gr}$ is anisotropic.  
A straight forward solution to Eq. (\ref{eq:bloch}) is to solve it
at grid points within a graphite unit cell 
using, for instance, a three-dimensional implicit method with
periodic boundary conditions at the edges of the cell.
The resulting six-dimensional function can be approximated as a series,
expanding around the diagonal elements, but this still gives a series of 
three-dimensional functions.  This greatly complicates Monte Carlo simulations
of the first-layer atoms using Eq. (\ref{eq:hitemp}), since storage 
requirements become
large and evaluating the density matrix by interpolating from three-dimensional
tables becomes excessively burdensome.  Thus, simulating the first
adsorbed layer using a high-temperature density matrix is a much more
complicated problem than simulating bulk helium.  One could always
avoid these problems by simply starting at a high enough temperature
so that the semiclassical 
approximation\cite{abraham87} can be used for atoms near the substrate, 
but then getting permutations accepted becomes exceedingly unlikely.

The problem becomes much simpler further above the substrate, where 
corrugations may be ignored.  The helium-graphite potential can be found
by laterally averaging over the surface, eliminating the x-y plane periodicity 
that complicates the solution near the substrate.  The helium atom
experiences only a z-dependent potential, so Eq. (\ref{eq:bloch}) can be
solved by separating $\rho_1^{Gr}({\bf r},{\bf r'},\tau)$ into x, y and z 
components.  The 
x and y components are one-dimensional, free-particle density matrices.  The
solution for $\rho(x,x';\tau)$, for instance, is
\begin{equation}
\label{eq:free1d}
\rho^{free}(x, x';\tau)=\lambda_t^{-1} 
	\exp[-\pi(x-x')^2/\lambda_t^2].
\end{equation}
A similar solution exists for $\rho(y,y';\tau)$.
The z-dependence is found by solving the parabolic partial
differential equation
\begin{equation}
\label{eq:zbloch}
\frac{\partial \rho}{\partial \tau}(z, z';\tau)
	=[(\hbar^2/2m) \partial^2/\partial z^2 - V^{Gr}(z)]\rho(z,z';\tau),
\end{equation}
where $V^{Gr}(z)$ is the laterally averaged potential.\cite{cole80}  
This can be solved by matrix squaring, or by an 
implicit method.\cite{numrep92}
The initial condition is that the density matrix is a delta function 
at $\tau=0$.  
We define the effective interaction for the 
helium-graphite density matrix, 
$U^{Gr}(z,z';\tau) \equiv -\ln[\rho(z,z';\tau)/\rho^{free}(z,z';\tau)]$.  
This is still a function of
two variables.  In order to make evaluating the density matrix
efficient during the Monte Carlo runs, we expand $U^{Gr}$ as a series of one
dimensional functions.  We rewrite $U^{Gr}(z,z')=U(\bar{z},\Delta z)$,
where $\bar{z} = (z + z')/2$ and $\Delta z = |z - z'|$.
The matrix is dominated by the diagonal
elements, so we expand it as a series about $(\Delta z)^2$:
\begin{eqnarray}
\label{eq:series}
U^{Gr}(z,z',\tau)=&&\frac{U^{Gr}(z,z,\tau)+U^{Gr}(z',z',\tau)}{2} \nonumber \\
	&+& \sum_{m} U_m(\bar{z}) (\Delta z)^{2m}. 
\end{eqnarray}

\begin{figure}[htp]
\epsfxsize=\figwidth\centerline{\epsffile{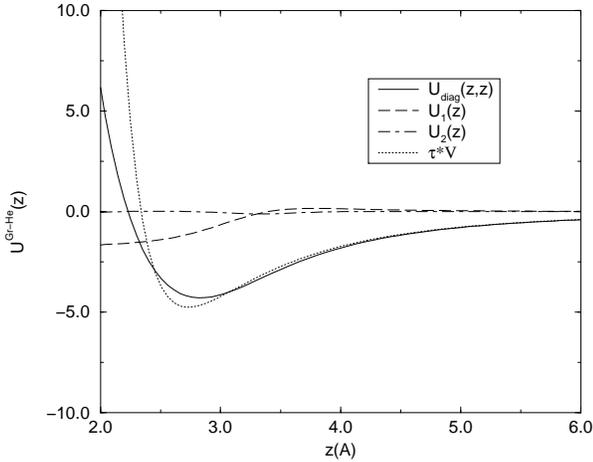}}
\caption{ The diagonal and lowest-order off-diagonal terms of the expansion
of $U^{Gr}$, Eq. (\ref{eq:series}).  The semiclassical approximation
is also shown.
The laterally averaged potential was used and $\tau=0.025 K^{-1}$.  
}
\label{fig:ugr}
\end{figure}

\begin{figure}[htp]
\epsfxsize=\figwidth\centerline{\epsffile{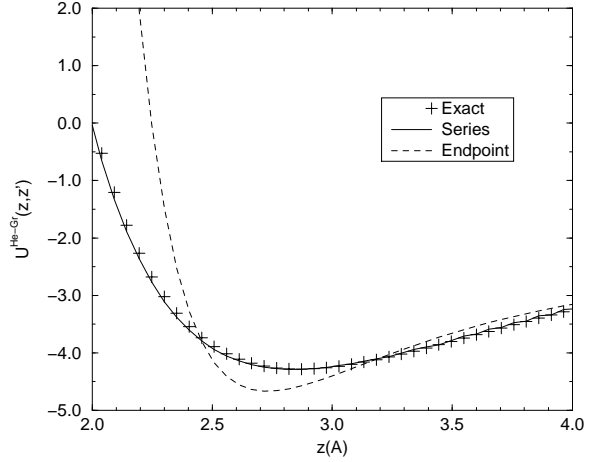}}
\caption{ The exact solution $U^{Gr}(z,z',\tau)$ 
for $z'=2.82\AA$ compared with the expansion, Eq. (\ref{eq:series}), and
the endpoint approximation using $\tau V$.
}
\label{fig:ugrfit}
\end{figure}

The average over the two diagonal parts of the solution in the first term
is called the endpoint approximation.  The functions $U_m(\bar{z})$ are 
found by $\chi^2$ fitting Eq. (\ref{eq:series}) to the exact solution. 
One simply terminates the series when the approximation is sufficiently
close to the exact solution.  Results for the diagonal solution
and the first two expansion terms are shown in Fig. \ref{fig:ugr}.  The
off-diagonal terms become negligible for $z> 4\AA$.  The diagonal solution
can be compared with the semiclassical approximation.  Figure \ref{fig:ugrfit}
compares the exact solution for off-diagonal elements to
the expansion, Eq. (\ref{eq:series}), and the endpoint approximation, 
$1/2[\tau V(z)+\tau V(z')]$.


\subsection{Sampling the density matrix}
With the first layer frozen, the
density matrix, Eq. (\ref{eq:hitemp}), for the active second layer
atoms can be written in the form
$\rho=\exp(-S)$, where 
\begin{eqnarray}
\label{eq:action}
S({\bf R},{\bf R}';\tau) \ = & &(3N_{act}/2) \ln(\lambda_t^2) 
		+\frac{\pi({\bf R}-{\bf R}')^2}{\lambda_t^2} \nonumber \\ 
	&+&\frac{1}{2}\sum_{i=1}^{N_{act}}\sum_{j=1}^{N_{act}} 
		U^{He}({\bf r}_{ij},{\bf r'}_{ij};\tau) \nonumber \\
	&+&\sum_{i=1}^{N_{act}}
		\sum_{j=1}^{N_{fr}}U^{He}({\bf r}_{ij},{\bf r}'_{ij};\tau) \nonumber \\
	&+&\sum_{i=1}^{N_{act}} U^{Gr}(z_i,z'_j;\tau),
\end{eqnarray}
where $r_{ij}=|{\bf r}_i-{\bf r}_j|$.
The number of active and frozen helium atoms is given by $N_{act}$ and
$N_{fr}$, respectively.  
In the polymer isomorphism, $S$ is the action for 
a system of interacting
polymers.  In sampling the paths, we are effectively choosing between
two different polymer configurations.  The one with the lower action
is the more favorable configuration, and is more likely to be chosen in
a Metropolis-style acceptance test.  

As in standard Monte Carlo simulations, the interaction $U^{He}$ is
cut off at some maximum distance $r_c \leq \min(L_x,L_y)$, where $L_x$ and
$L_y$ are the dimensions of the simulation cell.  
The long-range correction to the interaction felt by each particle is,
in cylindrical coordinates ($\rho$, z),
\begin{eqnarray}
\label{eq:ulr1}
U_{LR}^{He} = 2\pi\int_{0}^{\infty} n(z') dz' 
		\int_{\rho_c}^{\infty}\rho d\rho 
			 U^{He}(r,r;\tau),
\end{eqnarray}
where $r=\sqrt{\rho^2+(z-z')^2}$, 
$\rho_c^2 = r_c^2 -(z-z')^2$, and only diagonal elements need to 
be considered. The integral of $n(z')$ 
gives the density of the system.  We make
the approximation that the layer thicknesses can be treated as delta 
functions.  This is exact for the frozen first layer.  Then 
$n(z')=n_{fr}\delta(z'-z_{fr})+n_{act}\delta(z'-z_{act})$ and 
\begin{eqnarray}
\label{eq:ulr2}
U_{LR}^{He} = & &2\pi \int_{\rho_c}^{\infty} \rho d\rho 
	 [n_{fr} U^{He}(r,r;\tau) \nonumber \\
	&+& 1/2 n_{act} U^{He}(r,r;\tau)],
\end{eqnarray}
where $n_{fr}$ and $n_{act}$ are the densities of the first (frozen) 
and second (active) layers.
The factor of one-half before the contribution from the active layer
is needed to avoid double counting.
A similar long-range correction is added
to $\partial U^{He}/\partial \tau$ in the energy calculation.

As we have emphasized, particle permutations must be included in simulations
of superfluid helium.  These permutations 
correspond to splicing together two or more of
the polymer rings.  This splicing can be accomplished by proposing 
cyclic permutations involving one 
to four particle labels on inverse-temperature slice $i+n$ 
relative to slice $i$, where $n=2^l$ and {\em l} is the overall level
of the move.  The paths followed by the permuted particles on
the intermediate slices $i+1$ to $i+n-1$ that produce the permutation are then
filled in by successively bisecting the interval $i$ to $i+n$.
This is known as multilevel Monte Carlo sampling, an 
extension of the standard Metropolis method.  The interested reader is
referred to a recent review article on the subject.\cite{ceprev}

In our Monte Carlo runs for helium films, we take $l=3$, since this gives the
best balance between accepting single particle and multiple particle moves.
Increasing $l$ increases the number of permutations that can be accepted 
but decreases the overall acceptance rate, while decreasing $l$ has the
opposite effect.  The overall acceptance rate for the moves varies 
between 8\% and 15\%, depending on the density.  
Tests using $l=2$ at selected densities showed that the $l=3$ results
had converged.  The acceptance rate of multiparticle 
permutations is small, between 0.2\% and
0.3\% in the liquid phase.  We have found that similar small acceptance rates
are sufficient to obtain the superfluid density in bulk simulations.

\subsection{Calculating observables}
The expectation value of an observable, $A$, can be found from the trace, 
$<A>=Z^{-1} Tr A \rho$.  We use PIMC to calculate expectation values
for the total energy, the superfluid density, and the static structure factor. 
Below we give formulas for each of these calculations for a helium film
on a substrate.

The total energy is given by the expectation value
\begin{equation}
\label{eq:toteng}
E=\frac{3N_{act}}{2\tau} +<- \frac{\pi(\Delta {\bf R})^2}{\lambda_t^2 \tau} + 
\frac{dU^{He}_{total}}{d\tau} + \frac{dU^{Gr}_{total}}{d\tau}>.
\end{equation}
$\Delta \bf R$ is the change in the particle positions between two 
consecutive inverse-temperature slices.  The terms $U^{He}_{total}$
and $U^{Gr}_{total}$ are shorthand for the sums over the interaction
terms in Eq. (\ref{eq:action}).  Notice that the zero of the
total energy occurs at zero second-layer coverage, where there are no
active atoms. 

The superfluid density can be calculated using the winding number, $\bf W$,
for simulations that have periodic boundary conditions.
Nonzero winding numbers occur when particles, through a series of
permutations, are permuted with periodic images of themselves.  The winding
number is directly related to $\rho_s$, the superfluid density. 
\cite{ceprev}  For a system with periodic boundary conditions
in the x-y plane, the superfluid density is given by
\begin{equation}
\label{eq:winding}
\frac{\rho_s}{\rho}=\frac{m<({\bf W} \cdot {\bf L})^2>}{2 \beta \hbar^2 N_{act}},
\end{equation}
where the elements $L_x$ and $L_y$ are 
the dimensions of the simulation cell.  

Finally, structural information can be obtained with the static structure
factor,
\begin{equation}
\label{eq:ssf}
S({\bf k})= \frac{1}{N_{act}}<(\rho({\bf k})\rho(-{\bf k})>.
\end{equation}
We take $\hat{{\bf z}}$ to be perpendicular to the plane of the substrate, so
${\bf k} = (k_x,k_y)$.
$\rho({\bf k})=\sum_{i=1}^{N_{act}} \exp(i{\bf k} \cdot {\bf r}_i)$ is
the Fourier transform of the density.

\subsection{Testing the method}

As can be seen from the previous discussion, simulating helium
systems below the superfluid transition is an extremely complicated
task, and it is important to verify all parts of the method.
We have verified our
calculations for the solution to Eq. (\ref{eq:bloch_he}) by comparing
our results to published results for the Lennard-Jones\cite{cep84} potential
and to the Aziz potential.  
The solution to Eq. (\ref{eq:zbloch}) for the helium-graphite 
density matrix was checked by comparing the results obtained from the
matrix squaring and implicit solution methods.  We have verified that
the full Monte Carlo method outlined above works for bulk helium systems
by reproducing reported values for the energy, specific heat,
and superfluid density.\cite{cep86,runge92}  We believe these tests
sufficiently prove that our simulation method works and can be
extended to helium films.

\subsection{Choosing simulation cells}
We perform calculations with a variety of simulation cells that
are appropriate for examining different regions of the second-layer
phase diagram.
The first consideration is to choose a simulation cell that will
match the periodicity of the first-layer triangular solid.  
This can be done by using 
a rectangular unit cell with a two-point basis,  
with unit vectors ${\bf a}_1=a\hat{{\bf x}}$
and ${\bf a}_2 =\sqrt{3}a\hat{{\bf y}}$, where $a=3.015 \AA$.  
Two first-layer helium atoms are located in each unit cell at 
${\bf b}_1=0$ and ${\bf b}_2={\bf a}_1/2+{\bf a}_2/2$.
This gives
a coverage of 0.1270 atom/$\AA^2$, the fully compressed first-layer
density.\cite{greywall93}
In examining the second layer, our first goal is to scan the layer
at intermediate and higher densities by varying the number of particles and
to calculate the total energy at each density.  
For these calculations we use simulation cells with
dimensions ($5 {\bf a}_1, 3 {\bf a}_2$) and ($8 {\bf a}_1, 5 {\bf a}_2$), 
hereafter referred to as the
$5 \times 3$ cell and the $8 \times 5$ cell, respectively.  
The number of active particles in calculations using the 
$5 \times 3$ cell ranged from 8 to 21, corresponding to  
densities 0.1605 to 0.2159 atom/$\AA^2$.  Calculations with the $8 \times 5$
cell had 24 to 52 active particles, corresponding
to densities between 0.1651 and 0.2096 atom/$\AA^2$.  These two simulation
cells are nearly square, which is useful for calculating winding
numbers.  As will be discussed in Sec. \ref{sec:results}, the energy
calculations for the $5 \times 3$ cell are used to verify that
finite-size effects are not important in the $8 \times 5$ cell.  
Our conclusions about the coverage ranges of various phases
are drawn from results using the $8 \times 5$ cell.

\begin{figure}[htp]
\epsfxsize=\figwidth\centerline{\epsffile{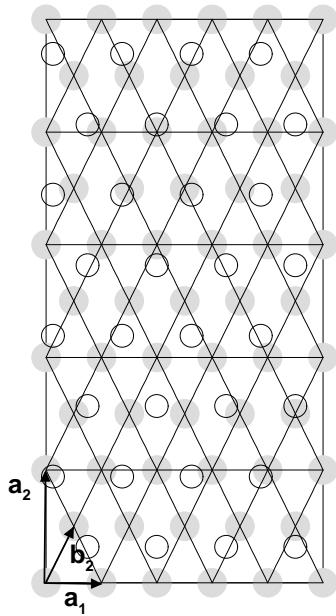}}
\caption{
Diagram of the $5 \times 5$ simulation cell.  The shaded circles denote
positions of the first layer atoms.  The 32 open circles denote
possible positions of atoms in the second-layer 
incommensurate triangular solid.  The arrows
indicate the unit vectors for the solid described in the text.  The lines
emphasize the triangular structure of the solid.
}
\label{fig:fig208}
\end{figure}

At high second-layer densities, commensurate and incommensurate triangular
solid phases occur.  In order to further 
investigate these phases, we use
simulation cells that can contain an integer number of unit cells of
both the first- and second-layer solids.  
That is, the simulation cells 
have the periodicity of both the first- and second-layer solids.  
It is also important to note that
the solid phases will tend to align with the x and y axes of the simulation
cell.  For the incommensurate solid we use a
cell with dimensions ($5 {\bf a}_1, 5 {\bf a}_2$), hereafter referred to as
the $5 \times 5$ cell.  This cell can accommodate 32 second layer atoms in
an equilateral triangular lattice.  A diagram of a second-layer
incommensurate solid in the $5 \times 5$ cell
is shown in Fig. \ref{fig:fig208}.  The second-layer solid is 
incommensurate with respect to the first since no supercell with
dimensions less than the minimum dimension of the simulation box can
be drawn in which both first- and second-layer atoms are periodically
repeated.

The simulation of 
the $\sqrt{7} \times \sqrt{7}$ 
triangular commensurate solid presents an additional problem since this 
structure is rotated with respect to the first layer.  This triangular
solid can be regarded as having a rectangular
unit cell with a fourteen point basis.  The unit vectors for this
solid are ${\bf s}_1=2{\bf a}_1+{\bf b}_2$ and 
${\bf s}_2=-2{\bf a}_1 + {\bf a}_2 + {\bf b}_2$.  
Note that $|{\bf s}_2| = \sqrt 3 |{\bf s}_1|$ and  
$|{\bf s}_i| = \sqrt{7}|{\bf a}_i|$, $i=1,2$.  We use simulation cells with 
dimensions ($2 {\bf s}_1,2{\bf s}_2$) and
($3 {\bf s}_1, 2 {\bf s}_2$) to identify the solid configuration 
and calculate the static structure factor.
The commensurate density 0.1996
atom/$\AA^2$ corresponds to 32 and 48 active particles, respectively, for
these two cells.  

\begin{figure}[htp]
\epsfxsize=\figwidth\centerline{\epsffile{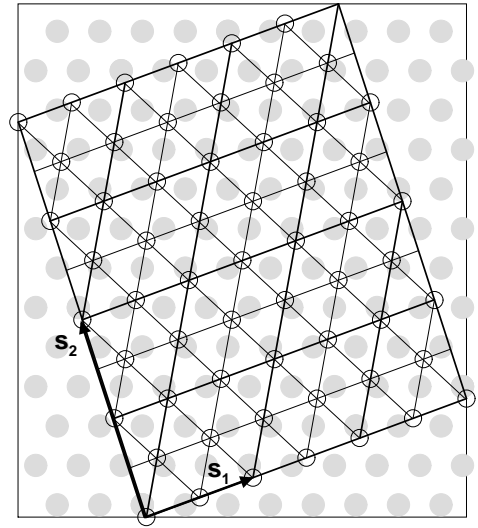}}
\caption{
Diagram of a simulation cell used for the $\sqrt 7 \times \sqrt 7$ solid.  
The dimensions are ($3{\bf s}_1,2{\bf s}_2$).
The shaded circles denote
positions of the first layer triangular solid.  The open circles denote
possible positions of the second layer registered solid.  The arrows
indicate the unit vectors for the solid described in the text.  The lines
emphasize the triangular structure of the solid.  The heavily shaded
lines indicate the $\sqrt 7 \times \sqrt 7$ supercells.
}
\label{fig:fig1996}
\end{figure}

A diagram of the ($3 {\bf s}_1, 2 {\bf s}_2$) simulation cell 
with the second layer
atoms in $\sqrt 7 \times \sqrt 7$ registry is shown in Fig. \ref{fig:fig1996}.
The large, rotated rectangle gives the bounds of the simulation cell.
First layer atom positions outside this rectangle are periodic images of
interior atoms.
Note that the location of the origin 
is arbitrary.  It is not necessary, for instance, to place it
at a high symmetry point of the first-layer lattice, such as over
a first-layer atom or at a potential minimum.  The essential requirements
for the existence of the partially registered solid are that once 
the origin is chosen, all of the two-dimensional space can be divided
up into periodically repeated superlattice unit cells (supercells), and 
that the relationships of the first- and second-layer atoms to each other
and to the supercell are the same in every supercell.  
We have chosen the
origin so that the second-layer atoms can be used to divide up the
rectangle into supercells.   These (primitive) supercells 
are the equilateral
parallelograms formed by the heavily shaded lines in the interior of
Fig. \ref{fig:fig1996}.  They can be seen to exactly fill the rectangle.
Second layer atoms are located at the four corners, 
on each of the four sides at the midpoints 
between the corners,
and at the center of each supercell, so there
is a four-point basis of second-layer atoms in each cell.  
The positions of the first-layer atoms can also be seen to be periodically
repeated in every supercell.  

\section{RESULTS}
\label{sec:results}

\subsection{Identification of phases}
Experimentally, there is evidence for liquid, commensurate solid and
incommensurate solid phases in the second layer.  We now describe the
identification of all three phases in our simulation.

\begin{figure}[htp]
\epsfxsize=\figwidth\centerline{\epsffile{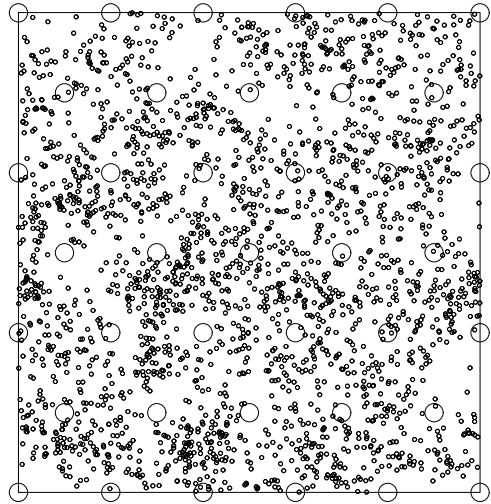}}
\caption{
Snapshot of a liquid configuration at 0.1778 atom/$\AA^2$, found using
the $5 \times 3$ simulation cell with twelve active particles and
$T=200$ mK.  Large circles indicate frozen first-layer atom sites.  The
instantaneous configuration of the second-layer atoms is indicated by the
small circles.
}
\label{fig:snap178}
\end{figure}

\begin{figure}[htp]
\epsfxsize=\figwidth\centerline{\epsffile{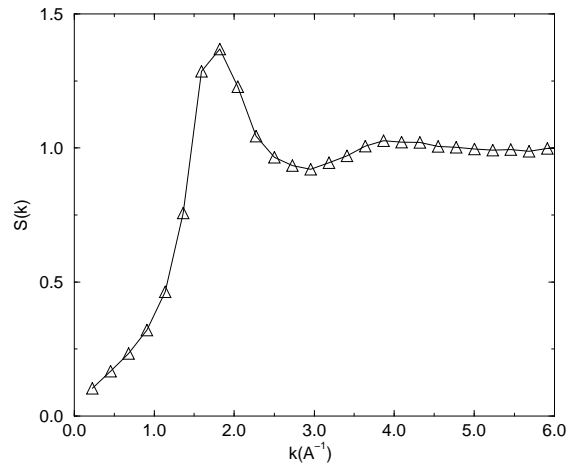}}
\caption{
The static structure function for the liquid phase 
at the density 0.1860 atom/$\AA^2$ and $T=500$ mK with 26 particles.  
}
\label{fig:ssf_liq}
\end{figure}

To find the liquid phase, we are guided first
by the torsional oscillator measurements, which detect a liquid phase 
between 0.174 and 0.187 atom/$\AA^2$.
We also find evidence that densities in this range are
liquid in our simulation.  Figure \ref{fig:snap178} shows a snapshot of
a typical liquid density.  The second-layer atoms obviously do not
possess spatial ordering, and the configuration covers the entire 
surface.
More direct evidence that the 
system has a liquid phase comes from that static structure factor.  
Figure \ref{fig:ssf_liq} shows the result of a calculation, which is 
typical of a self-bound liquid, at the coverage 0.1860 atom/$\AA^2$. 

Commensurate and incommensurate solid phases can be identified by 
a similar procedure.  A particularly nice feature of PIMC is that
these solids form on their
own, without any modifications to the high-temperature
density matrix, Eq. (\ref{eq:hitemp}).  In contrast, 
previous variational
calculations have used different trial wavefunctions for the liquid and solid
phases.\cite{whitlock88}  This can be avoided by using a shadow
wavefunction, but such calculations have not been performed for 
two-dimensional helium or helium films.

As demonstrated previously,\cite{pierce98} we have observed the 
$\sqrt 7 \times \sqrt 7$ commensurate solid phase in
our simulation for temperatures below 1 K.
The structure of this phase was determined
by examining snapshots of the configurations generated by the
simulation.  Particle paths of the second layer atoms were observed
to localize around the $\sqrt 7 \times \sqrt 7$
lattice sites shown in Fig. \ref{fig:fig1996}.
We note further that we do not bias the simulation of this solid by beginning 
the configuration at the commensurate lattice sites.  
The existence of the incommensurate solid, which occurs
at a higher density than the commensurate phase, has also been
demonstrated.  A snapshot
of this configuration generated by our simulation
can be found in our previous publication.\cite{pierce98}  
This phase matches the 
diagram shown in Fig. \ref{fig:fig208}.  We identify this phase as 
incommensurate because no supercell with dimensions
less than the minimum simulation box dimension can be drawn that
has both first- and second-layer atoms periodically repeated, in contrast
to the commensurate phase.

\begin{figure}[htp]
\epsfxsize=\figwidth\centerline{\epsffile{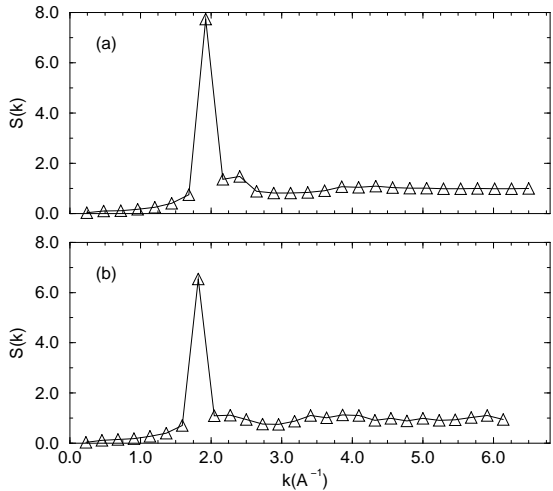}}
\caption{
The static structure factor calculated in the (01) direction for 
(a) the incommensurate solid at 
0.2083 atom/$\AA^2$ and 0.67 K with 32 particles, and
(b) the commensurate solid at 0.1996 atom/$\AA^2$ and 
0.50 K with 32 particles.
The errors are the size of the symbols. 
}
\label{fig:ssfk01_solid}
\end{figure}

The snapshots of the two solids 
are useful for visualizing their structure but are not actual tests for
their existence.  A direct measurement of correlation comes from
the static structure factor.  Results for
these calculations in the (01) reciprocal lattice
direction for the incommensurate and commensurate phases are 
shown Fig. \ref{fig:ssfk01_solid}(a) and (b).  The structure factor 
is normalized to $N_{act}$.  
The locations of these peaks give the correct lattice spacings for
the diagrams shown in the Figs. \ref{fig:fig208} and
\ref{fig:fig1996}.  The peak for the 
commensurate solid occurs at
1.82 $\AA^{-1}$, which gives the correct lattice constant, 
3.99 $\AA$, for the $\sqrt 7 \times \sqrt 7$ 
triangular solid.
Likewise, the peak for the incommensurate solid occurs at 1.93 $\AA^{-1}$,
corresponding to a lattice constant of 3.76 $\AA$, which is the 
correct lattice 
spacing for a triangular solid at 0.2083 atom/$\AA^2$.  

\subsection{T=0 phase diagram}

\begin{figure}[htp]
\epsfxsize=\figwidth\centerline{\epsffile{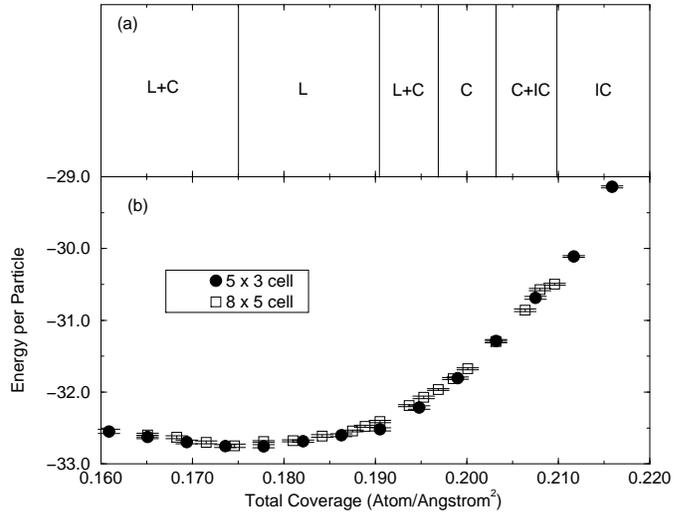}}
\caption{
(a) Summary of phase boundaries determined from applying the
Maxwell construction to the total energy of the $8 \times 5$ cell.  
The phases are liquid-gas (L+G), liquid (L), liquid-commensurate
solid (L+C), commensurate solid (C), commensurate solid-incommensurate solid
(C+IC), and commensurate solid (IC).
(b) The energy per particle for the $5 \times 3$ (circles) 
and $8 \times 5$ (squares) cells.
}
\label{fig:engcomb}
\end{figure}

Having identified the liquid, commensurate solid, and incommensurate solid
phases of the second layer, we now wish to find the boundaries for 
each of the phases.  We are able to identify the following density regions
at low temperature.
At low second-layer coverages, 0.1270 to 0.1750 atom/$\AA^2$, the system
is in a gas-liquid coexistence region, which consists of
a liquid droplet and a zero density gas.  The 
equilibrium density for the liquid is 0.1750 atom/$\AA^2$, and the 
layer is uniformly covered by a liquid phase from 0.1750 to 
0.1905 atom/$\AA^2$.  Above this density, the liquid
coexists with the $\sqrt 7 \times \sqrt 7$ commensurate solid phase 
discussed in the previous section.  This L-C coexistence occurs
from 0.1905 to 0.1970 atom/$\AA^2$, and is followed by the commensurate
phase between 0.1970 and 0.2032 atom/$\AA^2$.  The incommensurate solid 
phase begins to form above 0.2032 atom/$\AA^2$ and there is C-IC 
coexistence until 0.2096 atom/$\AA^2$.  Above this density, until layer
promotion to the third layer at 0.212 atom/$\AA^2$, the system is completely
in the incommensurate phase.  These results are summarized in 
Fig. \ref{fig:engcomb}(a).

Before discussing how these ranges were determined, we would first 
like to demonstrate that finite-size effects play an unimportant role
in the energy values used in the Maxwell construction.
Figure \ref{fig:engcomb}(b) shows the energy per particle found using
the $8 \times 5$ and $5 \times 3$ cells.  Almost all of the points 
calculated at similar densities in the two cells are consistent.  The primary 
``size effect'' is the limitation on the available densities
which may be examined for a given simulation cell.  

Phase ranges are determined by using the Maxwell double-tangent
construction, which identifies unstable regions associated with the
coexistence of two phases.  
A coexistence region at zero temperature in the thermodynamic
limit will have a total ground state energy that is the weighted average
of the two constituent phases' energy values.  
In Monte Carlo simulations the energy in the coexistence region will lie
above the coexistence line, either because the system remains in an unphysical
homogeneous state or because creating the phase boundary has a finite 
energy cost.\cite{allen87}  
We may thus identify a coexistence region
as the maximum range of densities in which all the intermediate
energy values lie on or above a line connecting the values at 
the endpoints.  We note that this version of the Maxwell construction is
somewhat different from other 
applications,\cite{whitlock88,saslow96,gordillo98}
which apply the Maxwell construction to the free energy dependence on
atomic area (inverse density).  Our method is appropriate for simulations
with constant area and varying particle number.

At finite temperatures, the Maxwell construction should be applied to
the total free energy.  Unfortunately, the free energy is not 
directly accessible from the PIMC simulation.  We instead make use of the
fact that at very low temperatures the free energy and the
energy are approximately the same, and become identical at zero temperature.
We can thus apply the Maxwell construction to low temperature energy
values to determine an effectively zero temperature phase diagram, provided 
that the values have converged to their zero temperature limits.  
To implement this procedure, we first
calculated energy values for a range of second-layer densities at 200 mK.
Selected energy values were recalculated at a higher temperature, typically
400 mK, and were seen to be within error bars of the 200 mK results.
This indicates
that our 200 mK calculations are effectively zero temperature results.

\begin{figure}[htp]
\epsfxsize=\figwidth\centerline{\epsffile{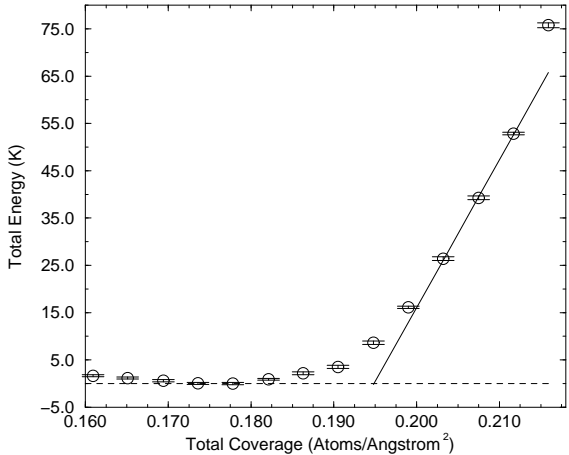}}
\caption{
The total energy found using the $5 \times 3$ simulation
cell with $N_{act}=8,9,\ldots,\ 21$ and $T=200$ mK.
The dashed line is gas-liquid coexistence line.  The solid line indicates 
a coexistence region terminating in an incommensurate solid phase.
}
\label{fig:smalleng}
\end{figure}

The application of the Maxwell construction to the total energy
values calculated using the $8 \times 5$ box has been shown
in our previous publication.\cite{pierce98} Figure \ref{fig:engcomb}(a)
summarizes the results.  
The energy minimum was determined to occur at 0.1746 atom/$\AA^2$ 
(30 particles).  
For comparison, Fig. \ref{fig:smalleng} 
shows the energy calculations using the $5 \times 3$ cell.  
These energy values have been shifted by $N_{act}e_{min}$ for
clarity, where $e_{min}=-32.754 \pm 0.020$ K. The energy minimum
occurs at 0.1778 atom/$\AA^2$ (12 particles).  Note that for both simulation
cells the minimum energy per particle occurs at nearly the same coverage value,
despite the
fact that the $8 \times 5$ cell is 2 2/3 times as large as the 
$5 \times 3$ cell.  In general, we find all the energy values calculated
with the two cells to be in agreement.  See Fig. \ref{fig:engcomb}(b).

The low density region of the second layer is known experimentally
to have coexistence between a gas phase and a superfluid liquid phase.
In order to determine the gas-liquid coverage range in our simulation,
we take the gas phase to have zero density at zero temperature and
thus zero total energy.  Two-dimensional calculations\cite{gordillo98} 
confirm that this assumption is correct for low temperatures.
A coexistence line can then be drawn between 
the beginning of the second layer, 
0.1270 atom/$\AA^{2}$, and the density
with the minimum energy per particle, which occurs between 
0.174 and 0.178 atom/$\AA^{2}$ in the $8 \times 5$ cell.  
The best $\chi^2$ 
parabolic fit to the energy data around the minimum gives
$\rho_0=0.1750(6)$ atom/$\AA^2$ for the density of minimum energy.  
The number in parenthesis is the error in the last digit.  
A similar coexistence line
can be identified in the $5 \times 3$ cell, Fig. \ref{fig:smalleng}.
We find that $\rho_0=0.1752(6)$, so finite-size effects on the 
liquid density are small.
At sufficiently
low temperatures, this liquid phase will become superfluid, as will be
discussed below.
All measured energy values for the
densities between 0.1270 atom/$\AA^2$ and $\rho_0$ lie above 
the coexistence line, so the system is  
in gas-liquid coexistence for this density range.  

The density of uniform liquid coverage, $\rho_0$, can be compared to 
experimental results.
For $T\leq 0.2$ K the second-layer heat
capacity measurements\cite{greywall91} show a probable gas-liquid
region roughly between 0.13 and 0.16 atom/$\AA^2$.  Within the
resolution available from the data, this phase can terminate anywhere 
from 0.1600 atom/$\AA^2$ up to, but not including, 0.1700 
atom/$\AA^2$ total coverage.
Since the first-layer coverage 
in the experiment is between
0.120 and 0.127 for these densities, gas-liquid coexistence terminates 
at second-layer coverages anywhere
from 0.033 to 0.050 atom/$\AA^2$.  For comparison, the gas-liquid phase
terminates at the second
layer coverage 0.0480(6) atom/$\AA^2$ in our simulation.  
Superfluidity is first observed in the torsional oscillator measurements 
at 0.174 atom/$\AA^2$.  Thus, the superfluid signal, 
as observed by this technique, becomes 
significant at the coverage
where our simulation determines that the second layer is 
uniformly covered by the liquid phase.

The density we determine for uniform liquid coverage can also be compared to
other simulations.  
In the two-dimensional calculations of Whitlock
et al., the equilibrium liquid 
coverage is 0.04356 atom/$\AA^2$ at zero temperature.
This result is supported by the low temperature results of 2D
PIMC calculations.\cite {gordillo98}  
This is slightly below our onset coverage, perhaps because we allow for 
particle motion perpendicular to the substrate.  Other calculations for
helium films also show a slight increase in the equilibrium density 
relative to the 2D result.  In the 
Monte Carlo calculation for the first layer of helium on 
graphite,\cite{whitlock98} the equilibrium density is determined to
be 0.0443 atom/$\AA^2$.  The effects of wavefunction spreading
will be even greater in the second helium layer.
Wagner and Ceperley's simulation of helium adsorbed on solid hydrogen
\cite{wagner94} also demonstrated that the liquid equilibrium density
increases when motion perpendicular to the substrate is allowed.  
They find a liquid coverage of 0.046(1) atom/$\AA^2$, comparable to
our result.  Thus the calculations of films with perpendicular spreading 
show a trend
toward higher liquid densities, with the onset density approaching the 
2D value as the helium-substrate potential becomes stronger.  From a 2D
viewpoint, this can be understood as a reduction of the hardcore 
repulsion, which allows for closer crowding.

At the highest second-layer densities, we can identify another
unstable region in the total energy values of the $8 \times 5$ cell
between 0.2032 and
0.2096 atom/$\AA^2$, corresponding to the C-IC mixed phase.  As shown
previously,\cite{pierce98} the coexistence line can be drawn
between the total energy values at these two densities.
The intermediate energy values lie on or above this line, so the region
has coexisting phases.  In particular, 
the energy value at 0.2080 atom/$\AA^2$ was found to lie
completely above the coexistence
line, providing an unambiguous signal for coexistence.
The range we find is in good agreement with the coexistence
region 0.2030 to 0.2080 atom/$\AA^2$ that can be determined from 
the heat capacity peaks.\cite{greywall93}
This phase coexistence is not a product of 
finite-size effects.  The beginning of a similar region may be identified 
between the densities 0.2032 and 0.2117 atom/$\AA^2$ in the 
$5 \times 3$ simulation cell, Fig. \ref{fig:smalleng}.
Phase coexistence in fact becomes
clearer in the $8 \times 5$ cell because we are able to examine
more density values in the unstable region.  

The presence of the C phase at 0.1996 atom/$\AA^2$ requires 
an L-C coexistence region between it and the liquid.
The region can also be identified in the $8 \times 5$ cell.
The endpoints of the L-C phase 
are 0.1905 and 0.1969 atom/$\AA^2$.  The intermediate energy values lie
on the coexistence line within error bars.
The L-C range is in reasonable agreement 
with the coexistence 
range 0.1871 to 0.1970 atom/$\AA^2$ determined from
heat capacity measurements.\cite{greywall93}
Torsional oscillator measurements\cite{reppy93} also indicate that 
the coexistence region begins at about 0.187 atom/$\AA^2$.
The L-C phase cannot be determined in the $5 \times 3$ cell 
due to the coarseness of the coverage grid.  

\subsection{Other properties}

\begin{figure}[htp]
\epsfxsize=\figwidth\centerline{\epsffile{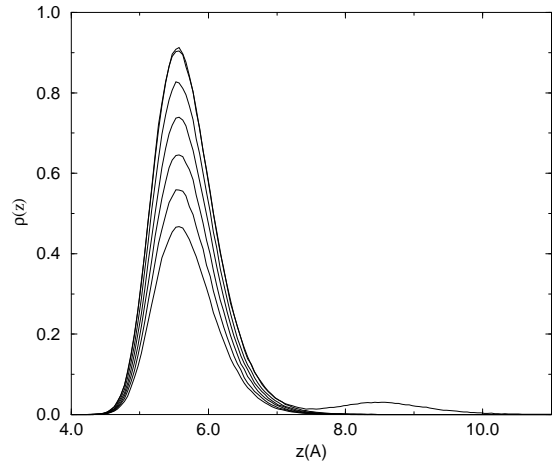}}
\caption{
Density profiles for the second layer found using the $5 \times 3$
cell, with densities 0.1694, 0.1778, 0.1863, 0.1948, 0.2032, 0.2117,
and 0.2159 atom/$\AA^2$.
}
\label{fig:zdist}
\end{figure}

Figure \ref{fig:zdist} depicts the density profiles for selected 
layer densities.  These
plots are normalized such that integrating $\rho(z)$ gives the number
of particles.  Promotion to the third layer can be 
clearly observed at the highest density shown, 0.2159 atom/$\AA^2$, so
we conclude that layer promotion occurs between 0.2117 and 0.2159 atom/$\AA^2$.
This is in excellent agreement with the completion density 
of 0.212 atom/$\AA^2$
determined from the heat capacity measurements.\cite{greywall91,greywall93}
A somewhat lower value of 0.204 atom/$\AA^2$ for third layer
promotion is obtained from the isothermal 
compressibility minima.\cite{zimmerli92,reppy96}
Also of note, Whitlock et al. \cite{whitlock98} estimate that promotion
to the third layer begins at the second-layer coverage of 0.08 atom/$\AA^2$,
quite close to but somewhat lower than our value of 0.085 atom/$\AA^2$.  

The temperature dependence of the energy and superfluid density 
at a sample liquid 
density of 0.1778 atom/$\AA^{2}$ have been determined.  This coverage
corresponds
to a second-layer coverage of 0.0508 atom/$\AA^2$.
Values were calculated using the $5 \times 3$ simulation cell
with twelve active particles, and are illustrated in a previous
publication.\cite{pierce98}
The superfluid density is relative to the second-layer density.
Both the energy and the superfluid density converge to 
the ground state for temperatures
below 0.8 K.  The slow decay of the superfluid density at higher 
temperatures is a typical 2D finite-size effect.\cite{norbert2d}
The superfluid density values have been $\chi^2$ fit to the 
solution to the Kosterlitz-Thouless (KT) recursion 
relations.\cite{nelson77}  From the intersection of the
fit and the KT transition line, we estimate
the transition temperature to be $T_c \approx 0.88 K$.  For comparison, 
the 2D PIMC simulation\cite{gordillo98} obtains $T_c=0.86 \pm 0.02 K$
at 0.0508 atom/$\AA^2$.  

The specific heat of the liquid, commensurate solid, and
incommensurate solid phases can be found 
by differencing the energy per particle 
with respect to temperature.  This was shown 
in our previous publication.\cite{pierce98}  For the liquid phase, a
broad, low peak with a maximum value at 1.18 K was found.  This is comparable
to the experimental heat capacity results,\cite{greywall93} which have a peak
at 1 K.  For the commensurate solid phase, a  specific heat peak at about 1.5 K
was found.  This is comparable to the heat capacity 
measurements\cite{greywall93} at 
similar density values, which also have 
a peak at 1.5 K.  This close agreement provides some additional evidence that
the $\sqrt 7 \times \sqrt 7$ C phase occurs in the experiment.  
Finally, for the IC solid, a peak at 0.7 K was obtained, somewhat lower 
than the peak in
the heat capacity measurements at the same density, which occurs at 1 K.  

\section{SUMMARY}
\label{sec:conclusion}
A number of recent experiments indicate that the second layer of helium
on graphite has an interesting phase diagram.  Torsional oscillator
measurements detect superfluidity over a narrow density range in this 
layer.\cite{reppy93,reppy96} 
Neutron scattering\cite{carneiro81,lauter91,lauter92} detects  
an incommensurate solid phase at high densities.
Heat capacity measurements\cite{greywall91,greywall93}
have found evidence for liquid-gas coexistence and the incommensurate
solid phase.  The heat capacity data also show the existence of
an intermediate phase between the liquid and incommensurate solid, which
is possibly a commensurate solid.  The existence
of this commensurate solid phase would explain the disappearance of 
superfluidity at higher second layer coverages.  Motivated by these
experiments, we have undertaken a simulation of this layer.

In order to study the second layer with Monte Carlo for a range of 
temperatures, it is necessary to develop a method that
incorporates both particle permutations and the effects of the substrate
and the solid first layer on the second.  Permutations are necessary to obtain
the superfluid phase.  The effects on the solid first layer must be included
since the commensurate second layer solid is partially
registered with respect to the first layer.  First layer and substrate effects
also play a role in the formation of the incommensurate solid phase, 
which replaces the commensurate phase before layer promotion begins.

We have developed a path integral Monte Carlo method that includes the 
above features.  Particle permutations were
included in the simulation using a method developed for bulk 
helium\cite{ceprev}.
We have shown that the helium-helium and helium-graphite
interactions can be incorporated into the simulation by using effective 
interactions found from the exact 
solutions for the interacting part of the appropriate density matrices.  
Realistic helium-helium and helium-graphite potentials are used to find
these effective interactions.
For the helium-graphite effective interactions, we have shown 
how this solution may be approximated so that 
off-diagonal matrix elements may be efficiently and accurately included
in Monte Carlo sampling.  The interaction of the second layer of helium 
atoms with the solid first layer were approximated by placing first 
layer atoms at triangular lattice sites with
a lattice spacing that gives the completed first layer density.  These
atoms were located at a fixed height above the substrate, given by 
the minimum of the effective helium-graphite 
interaction.  Configurations of 
these atoms were not sampled, which allowed us to scan second layer
densities with a finer grid.  Therefore, we study the second layer atoms 
under the influence of their mutual interactions and a static potential 
produced by the frozen graphite substrate and the frozen first layer
helium atoms. This  approach 
ignores effects on the second layer from the  zero point motion of the 
first layer solid and first layer compression effects. We 
feel this is a reasonable approximation because the relatively
high Debye temperature of the completed first layer\cite{dash83} means
that it will be relatively stiff for the temperatures of our simulation. 
Compression effects on the first layer by the second are
most important for low second layer
densities\cite{bretz78}, below the range of our simulation.

Using the above simulation method, we were able to identify, in order
of increasing density, superfluid
liquid, $\sqrt 7 \times \sqrt 7$ commensurate triangular solid, and 
incommensurate triangular solid phases from particle 
configurations and static structure factor calculations.  
We also calculated the specific heat for each of these phases and observed
peaks in general agreement with experiment.

The density ranges at effectively zero temperature of each of the second layer 
phases and their coexistence regions were determined using the Maxwell 
construction.  We found that at
low densities, the layer is phase separated into a liquid droplet 
and a zero density
gas.  The range of this phase is 0.1270 to
0.1750 atom/$\AA^2$.  Gas-liquid coexistence ends at the equilibrium 
density for the liquid phase.  This occurs at 0.1750 atom/$\AA^2$, which is
the density with the minimum energy per particle.
This density was found to be insensitive to finite-size effects, 
and is in excellent agreement with the onset of 
superfluidity determined by torsional oscillator measurements.  
It is also consistent with heat capacity measurements.
We demonstrated that the liquid phase in our simulation is superfluid, 
and we determined that
the transition temperature was close to the value determined for a purely
2D superfluid at the same density.  

The helium layer is uniformly covered in our simulation by the liquid phase 
from 0.1750 to 0.1905 atom/$\AA^2$, at which point liquid-commensurate solid
phase coexistence begins.  The onset of this coexistence terminates 
superfluidity, since the growth of the solid phase disrupts 
the connectivity required to detect
the superfluid.  Experimentally, liquid-commensurate solid phase
coexistence has been determined to begin at 0.1870 atom/$\AA^2$ by both
torsional oscillator and heat capacity measurements.
We determined that the liquid phase is completely replaced by the
$\sqrt 7 \times \sqrt 7$ commensurate solid for densities above 
0.1970 atom/$\AA^2$, in good agreement
with heat capacity measurements.  Phase coexistence between the commensurate
and incommensurate solid phases
begins at 0.2032 atom/$\AA^2$.  For coverages above 0.2080 atom/$\AA^2$, 
the incommensurate
solid is the only phase occurring until layer promotion.  These ranges 
for the solid coexistence and the incommensurate solid are
in agreement with the heat capacity measurements.  The density ranges
for all the second layer phases described above are
summarized in Fig. \ref{fig:engcomb}(a).  Finally, we 
observed layer promotion for coverages above 0.2117 atom/$\AA^2$, in 
excellent agreement with experiment.

\section{ACKNOWLEDGMENTS}
This work was supported in part by the National Aeronautics and Space
Administration under grant number NAG3-1841. 
Some of the calculations for this work were performed using the computational
facilities of the Supercomputer Computations Research Institute and
the National High Magnetic Field Laboratory at the Florida State University.
We wish to thank W. Magro and M. Boninsegni for allowing us to compare
outputs of our program to theirs for the helium-helium high-temperature 
density matrix so that we could verify our solution.  
M. P. wishes to thank M. C. Gordillo for discussions on the superfluid 
transition temperature.

\end{document}